\documentclass[twocolumn,epsfig,pre]{revtex4}

\usepackage{graphics,graphicx,epstopdf}
\usepackage{amssymb,amsmath,bm}
\usepackage{booktabs}
\usepackage{hyperref}
\usepackage{latexsym,color}

\begin{document}

\renewcommand{\thefootnote}{\fnsymbol{footnote}}
\renewcommand{\theequation}{\arabic{section}.\arabic{equation}}

\title{Tracer dynamics in dense soft-colloidal suspensions: From free diffusion to hopping}


\author{Jimpaul Samukcham}
\author{Lenin S.~Shagolsem}
\email{slenin2001@gmail.com}
\affiliation{Department of Physics, National Institute of Technology Manipur, Imphal, India} 


\begin{abstract}
Tracking of individual particle and studying their motion serves as a direct means to understand the dynamics in crowded and complex environments. In this study, the dynamics of tracer particles in the matrix of dense soft-colloidal suspensions in fluid phase is studied by means of dissipative particle dynamics simulations. By considering relatively large tracer (three times that of colloid) we systematically explore the interplay between the environment in which the tracer undergoes motion and interaction with the environment on the dynamics for temperatures close to the thermodynamic freezing transition where the effect of pair-wise interaction is significant compared to thermal energy. To this end we consider three fluid systems differing in the degree of softness (i.e., ultra-soft, intermediate, and hard) of the constituent colloidal particles, also change tracer types in the sense that we vary the degree of softness of the tracer w.r.t. colloids from ultra-soft to very hard. It is found that for tracer in ultra-soft colloidal fluid, at long times, the motion is diffusive for all tracer types, however the relaxation time (or diffusion constant) increases (or decreases) with increasing hardness of tracer at a given temperature. Interestingly, for tracer in hard colloidal fluid, the motion changes from a free diffusion (continuous trajectory) to that of hopping where there is intermittent jumps following a long period of localized vibrations and consequently displacement distribution function show higher order peaks indicating different dynamics at different time (or length) scales. 
\end{abstract}

\maketitle


\section{Introduction} \label{sec: intro}

Unlike hard sphere colloids soft-colloids such as star polymers, dendrimers, nano/micro gels, emulsion droplets, block-copolymer micelles, polymer-grafted nanoparticles are permeable, deformable, can have network-like  architecture, and it can respond to different solvent conditions. \cite{stieger2004,karg2006,hatto2000,karg2011,nayak2005,karg2019} 
In the recent past, soft colloidal systems have been the subject of intense research addressing both equilibrium properties (with major focus on the structural change due to chemical composition and external stimuli like temperature, solvent quality and ionic strength) and non-equilibrium properties. Because of its unique properties suspension of soft-colloids have huge potential applications e.g. in paints, pharmaceutical as biocatalyst, biological sensor, drug delivery, etc.~to mention a few. \cite{karg2019,granick2003}  \\

Soft colloids show non-monotonic density dependence of the freezing transition~\cite{stillinger1976,louis2000,lang2000} with a complex crystalline structures,\cite{likos2002,pamies2009} a shift of the glass line to higher effective volume fractions with increasing softness,\cite{heyes2009,gonzalez2009} and a density dependent structural ordering and phase behaviour. \cite{berthier2012,paloli2013}  
Furthermore, in a recent study by Scotti and coworkers~\cite{scotti2018} the response of hollow microgels with respect to regular ones in overcrowded environments is investigated and found that hollow cavity allows for an alternative mechanism to respond to the compression, i.e., the polymer chains are pushed and rearrange within the cavity. This clearly illustrates that the response of soft colloids is sensitive to the micro-structure as well. 
Soft colloidal system also display rich dynamical features, e.g., in a dense suspension of responsive soft colloids a strong influence of solvent-mediated core swelling on the concentration dependence of self-diffusion is observed.\cite{loppinet2005} Mechanical response of the system is also affected by the particle softness, where, for instance, increase of shear modulus as a function of increasing volume fraction is relatively weak in the case of softer particles.\cite{grande2008} See the work of Likos for a brief summary of equilibrium and dynamical properties of concentrated solutions of star polymers and dendrimers.\cite{likos2006} \\   

On the other hand, a dense suspension of soft colloidal system is realized in living systems as well, where inside the nucleus of eukaryotic cells or prokaryotic cell proteins or macromolecules of different shape and sizes are densly packed. Understanding the dynamic organization of cellular components, mechanical response to external stimuli of such highly complex systems is an area of active research.\cite{hameed2012,manzo2015,barkai2012,izeddin2014} In this regards, tracking of individual particles and studying their motion both experimentally (under a microscope) and theoretically (through computer simulations) serves as a simple and direct means to probe the dynamics in crowded and heterogeneous complex bio-environments.\cite{manzo2015,barkai2012,izeddin2014}  
Similar work on the mobility of tracer (or nanoparticles) in a polymer matrix with binding zones, and the motion of polymer tethered nanoparticles in unentangled polymer melts have been reported whose results is important in designing the drug carriers moving through cells and extracellular matrices.\cite{sumanta2016,ge2020,ge2019,schuster2015} \\

In this work, motivated by the importance of particle tracking experiments in revealing the physics of the complex fluid environment, we consider a relatively simple system of dense soft-colloidal suspension (in fluid-phase) with varying degree of softness and study the dynamics of tracers embedded in such fluid systems. (In the following, we refer colloids as fluid particles in general.) 
To the best of our knowledge, despite having numerous reports on the particle tracking experiments and theory, the role of having soft or hard tracer w.r.t. the medium colloidal particles has not been addressed adequately. Therefore, in this study, by means of computer simulations we systematically explore the effect on the tracer dynamics (and the corresponding statistics of particle trajectories) when the tracer is hard or soft with respect to the constituent colloid particles of the fluid. In particular, for a fixed degree of softness of fluid particles, the tracer-fluid interaction is varied and study its consequences on the tracer dynamics in the vicinity of thermodynamic melting or freezing temperature of the fluid where the effect of pair-wise interaction becomes significant and thus the difference between the systems get amplified. \\

The remainder of the paper is organized as follows. In section~\ref{sec: model-description}, the model and simulation details are presented and in section~\ref{sec: fluid-summary}, following a brief discussion on the thermodynamics and particle dynamics of the fluid, we study the dynamics of tracer under systematic variation of softness parameters and finally conclude the paper in section~\ref{sec: conclusion}. 
\section{Model and simulation details}
\label{sec: model-description}

In order to study the tracer dynamics in model dense soft-colloidal suspensions we adopt dissipative particle dynamics (DPD) simulations, \cite{espanol1995,groot1997,allen-book,frenkel-book} where the method allow modelling of large macromolecules (e.g., polymers and colloids) over large length and time scales. In this model, the total force ${\bf f}_i$ acting on a particle $i$ due to particle $j$ is given by 
\begin{eqnarray}
{\bf f}_i = {\bf f}_{\rm C}(r_{ij}) + {\bf f}_{\rm D}(r_{ij},v_{ij}) + {\bf f}_{\rm R}(r_{ij})~,
\end{eqnarray}
where ${\bf f}_{\rm C}$, ${\bf f}_{\rm D}$, and ${\bf f}_{\rm R}$ are conservative, dissipative, and random forces, respectively. Here, $r_{ij}=|r_i-r_j|$ and $v_{ij}=v_i-v_j$. In this model, the conservative force, a soft repulsive force, is given by 
\begin{eqnarray}
{\bf f}_{\rm C}(r_{ij}) = \left\{
\begin{array}{l l}
a_{ij}\left(1-\frac{r_{ij}}{r_c}\right){\bf e}_{ij}~, & \quad r_{ij}<r_c \\ 
0~, & \quad r_{ij}\geq r_c
\end{array} \right. .
\label{eqn: force-conservative}
\end{eqnarray} 
Here, the directional vector ${\bf e}_{ij}=({\bf r}_i-{\bf r}_j)/r_{ij}$, and $r_c$ is the interaction cut-off distance. The energy due to ${\bf f}_{\rm C}$ is shifted to zero at $r_c$. The strength of repulsion between a pair of particles $i-j$ is controlled via parameter $a_{ij}$, where large (or small) value of $a_{ij}$ would correspond to hard (or soft) particle. The dissipative and random forces are given by 
\begin{eqnarray}
{\bf f}_{\rm D}(r_{ij},v_{ij}) &=& -\gamma_{ij}\omega_{\rm D}(r_{ij})\left( {\bf e}_{ij}\cdot{\bf v}_{ij} \right){\bf e}_{ij}~,~~{\rm and} \label{eqn: fd} \\
{\bf f}_{\rm R}(r_{ij}) &=& \xi_{ij}\omega_{\rm R}(r_{ij})\theta_{ij} {\bf e}_{ij}~, 
\label{eqn: fr}
\end{eqnarray}
respectively, with $\gamma_{ij}$ the fricton coefficient, $\xi_{ij}$ amplitude of the noise, $\theta_{ij}(=\theta_{ji})$ the Gaussian white-noise, and $\omega_{\rm D}$ and $\omega_{\rm R}$ are weight functions which vanishes for $r\ge r_c$. The above functional form of ${\bf f}_{\rm D}$ and ${\bf f}_{\rm R}$ ensure that the system conserves momentum and hence correctly recover the hydrodynamic behavior on sufficiently large length and time scales.\cite{allen-book,frenkel-book} Between the weight functions, and between the amplitude of the noise and $k_BT$ the following relations 
\begin{eqnarray}
\omega_{\rm R}(r_{ij}) = \omega_{\rm D}^{1/2}~,{\rm and}~~~
\xi_{ij} = (2\gamma_{ij} k_B T)^{1/2}
\end{eqnarray}
must be obeyed to satisfy the fluctuation-dissipation theorem. We take the following form of the weight function
\begin{eqnarray}
\omega_{\rm D} = (\omega_{\rm R})^2 = \left\{
\begin{array}{l l}
\left(1-\frac{r_{ij}}{r_c}\right)^2~, & \quad r_{ij}<r_c \\ 
~0~, & \quad r_{ij}\geq r_c
\end{array} \right. .
\end{eqnarray}
All the quantities used are in reduced units, where energy is measured in units of $k_BT$, time in units of $t_0=r_c(m/k_BT)^{1/2}$, and we set $\xi_{ij}=3k_BTt_0^{1/2}/r_c$ and $\gamma_{ij}=4.5k_BTt_0/r_c^2$.\cite{allen-book,frenkel-book} \\

\begin{figure*}[]
	\begin{center}
		\includegraphics[width=0.8\textwidth]{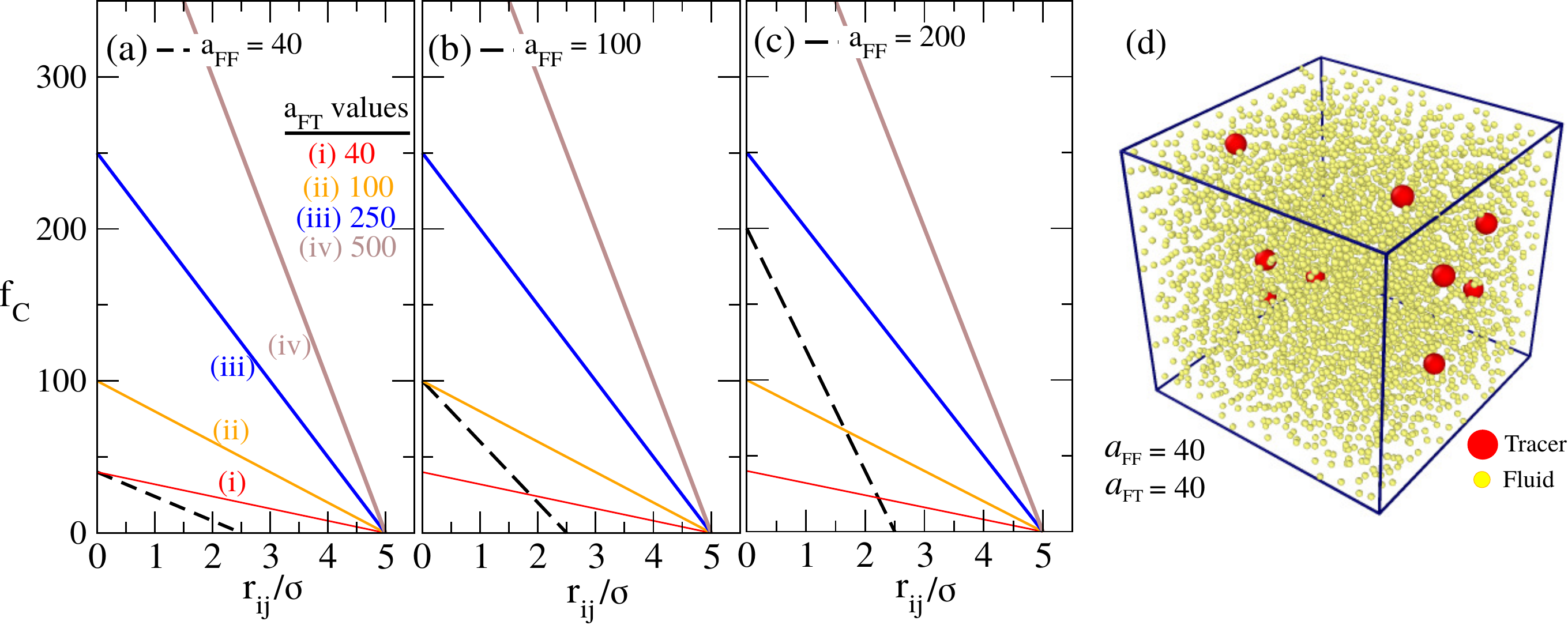} 
		\caption{Conservative force ${\bf f}_{\rm C}$ (eqn.~\ref{eqn: force-conservative}) between a pair of fluid particles (dashed line) and fluid-tracer (solid line) at fluid-fluid softness parameter $a_{\rm FF}=$ (a) 40, (b) 100, and (c) 200 plotted together with that of fluid-tracer at different values of fluid-tracer softness parameter $a_{\rm FT}$ as indicated in the figure. (d) Typical equilibrium configuration of the fluid+tracer system at $T=4$ (fluid phase) and $a_{\rm FF} = a_{\rm FT} = 40$. As indicated in the figure, tracer (fluid) particles are represented by red (yellow) spheres.} 
		\label{fig: snapshot-system}
	\end{center}
\end{figure*}

The concentrated colloidal fluid system is modeled by considering $N_0=4000$ DPD particles confined in a cubic box of dimension $L=19\sigma$, with $\sigma$ effective particle diameter. The simulation box is periodic in all dimensions. The reduce number density of the system is $\rho=0.583~(=N_0/{\rm Vol.})$ and the corresponding packing fraction $\eta \approx 0.3~(=\pi\rho\sigma^3/6)$. The fluid particle diameter and mass are set to unity, and the value of interaction cutoff distance $r_c=2.5\sigma_{ij}$, with $\sigma_{ij}=(\sigma_i+\sigma_j)/2$ the effective diameter between particle $i$ and $j$. To this we add $N_T=10$ tracer particles of diameter $\sigma_T=3\sigma$ (and thus it occupies a volume fraction of $\eta\approx 0.02$). Mass of the tracer particle is also fixed at unity, and tracer-tracer/fluid interaction cutoff distance is fixed at $2.5\sigma_{ij}$. Thus, for the fluid-fluid interaction $r_c=2.5\sigma$ and that of the fluid-tracer interaction $r_c=5\sigma$. 

\begin{table}[h]
\caption{Repulsion parameters $a_{ij}$ between a pair of particles $i-j$ and the corresponding nature of interaction in terms of degree of softness of colloid/tracer.} 
\centering
\begin{tabular}{@{}lll@{}} 
	\hline
	$i-j$ & $a_{ij}$ & {Nature}  \\
	\hline
	Fluid--Fluid   & 40   & ultra-soft \\
				   & 100  & intermediate \\
	   			   & 200  & Hard \\
	Tracer-Fluid   & 40 -- 500   & ultra-soft -- very hard   \\
	Tracer-Tracer  & 200	& intermediate \\ 
	\hline 
\end{tabular}
\label{table: parameters}
\end{table}

We consider three types of fluid systems differing in their degree of softness, characterized by the value of fluid-fluid (FF) repulsion parameter $a_{\rm FF}$ (see equation~\ref{eqn: force-conservative}), of the constituent particles: $a_{\rm FF}=40,~100,~200$ (ultra-soft to relatively hard colloids) to which tracer particles are added.  
In these fluid systems, the fluid-tracer (FT) interaction parameter $a_{\rm FT}$ varies in the range $40-500$. The interaction parameters used in the simulations are summarized in table~\ref{table: parameters}. After preparing the initial configurations the systems are first relaxed at high temperature ($T=5$) at which the system is in disordered state and then cooled down to the desired value of $T$ followed by relaxation and production runs. Typical relaxation and production runs are of $1\times 10^6$ and $1\times 10^7$ simulation steps, respectively, with integration timestep of 0.01. All the results reported here are for the equilibrated systems in the sense that ensemble averaged properties (e.g., mean potential energy per particle) are time-independent. For simplicity we keep the friction coefficient (see equation~\ref{eqn: fd}) to be same for both fluid and tracer particles in our simulations. \\

A typical equilibrated sample in the fluid phase obtained from DPD simulations is shown in figure~\ref{fig: snapshot-system} along with the conservative force to highlight the strength of the forces between different pairs of particles. To study the tracer dynamics we restrict the system close to the thermodynamic freezing/melting temperature $T^*$, where the system is sensitive to the applied potential and kinetic energy dominance is less. As described in section~\ref{subsec: fluid-thermodynamics}, the value of $T^*$ for the different fluid systems are estimated through continuous heating/cooling thermal cycles. Note that adding a very small amount of tracer particles has no significant effect on the thermodynamics of the reference fluid. 
All the simulations are carried out at constant NVT ensemble~\cite{allen-book,frenkel-book} using LAMMPS code.\cite{lammps} 


\section{Results and discussion} \label{sec: fluid-summary} 


\subsection{Statics and dynamics of the dense colloidal suspension: A brief summary}
\label{subsec: fluid-thermodynamics} 

In order to locate $T^*$ for the reference fluid systems, the relaxed samples at $T=5$ are then cooled down continuously to $T=0.1$ using $1\times 10^7$ simulation steps followed by heating up continuously back to $T=5$ at the same rate. During the continuous heating/cooling process we monitor the mean energy per particle, $\langle u \rangle$, of the system and observe an abrupt change at a particular value of $T$ which is different for heating and cooling curves, and thus a hysteresis loop is present during a complete cooling-heating cycle. \\

\begin{figure}[]
	\begin{center}
		\includegraphics[width=0.4\textwidth]{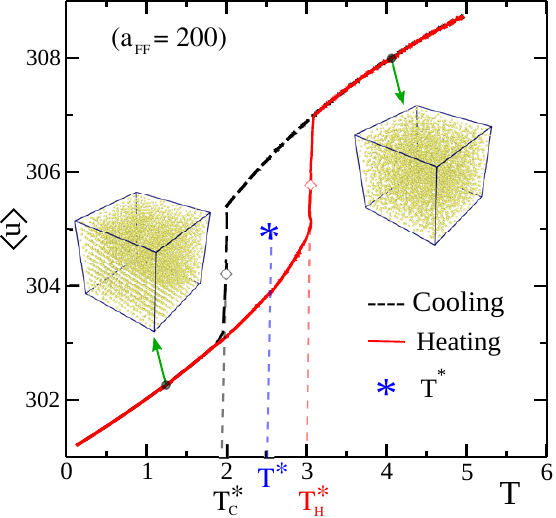} 
		\caption{Change of mean potential energy per particle $\langle u \rangle$ of the reference fluid system with $a_{\rm FF} = 200$ during continuous cooling and heating. $T_{\rm C}^*$ and $T_{\rm H}^*$ are the positions at which $\langle u \rangle$ changes abruptly during cooling and heating, respectively, and $T^*$ is the thermodynamic melting temperature (equation~\ref{eqn: transition-T}). Inset: simulation snapshots of the system at the indicated temperatures.} 
		\label{fig: HC-cycle-a200} 
	\end{center}
\end{figure}
Thermal hysteresis loop for the reference colloidal fluid with $a_{\rm FF}=200$ is shown in figure~\ref{fig: HC-cycle-a200} along with the representative configuration at high (fluid phase) and low (ordered phase) temperatures. Similar thermal hysteresis loops are obtained for different fluid systems. In the presence of thermal hysteresis the value of $T^*$ is approximately given by 
\begin{eqnarray}
T^* = T_{\rm C}^* + T_{\rm H}^* - \sqrt{T_{\rm C}^* T_{\rm H}^*}~,
\label{eqn: transition-T}
\end{eqnarray}
where $T_{\rm C}^*$ and $T_{\rm H}^*$ are the transition temperatures on cooling and heating (see figure~\ref{fig: HC-cycle-a200}), respectively.\cite{luo2004} The value of $T^*$ thus obtained for the fluid systems are $T^*\approx 0.5,~1.24,~{\rm and}~2.51$ for $a_{\rm FF}=40,~100,~{\rm and}~200$, respectively. Since the systems have different $T^*$ values comparison among them is done at the same value of reduced temperature $\delta$ defined as $\delta\equiv (T-T^*)/T^*$. We consider $\delta$ in the range $-0.1 \le \delta \le 0.3$ and this, e.g., for $a_{\rm FF}=200$ system (see figure~\ref{fig: HC-cycle-a200}) would be in the temperature range $2.26 \le T \le 3.26$ which is in the vicinity of $T^*$ or falls largely in the hysteresis loop region. Also note that the lowest temperature we consider, i.e., $\delta=-0.1$ or $T\approx 2.26$ ($a_{\rm FF}=200$ system) is larger than $T_{\rm C}^*$, but smaller than $T_{\rm H}^*$. Therefore, the system is expected to be still in the fluid/disordered state even at $\delta=-0.1$ during cooling, whereas it is expected to be in the ordered state when heating up. \\
\begin{figure}[]
	\begin{center}
		\includegraphics[width=0.45\textwidth]{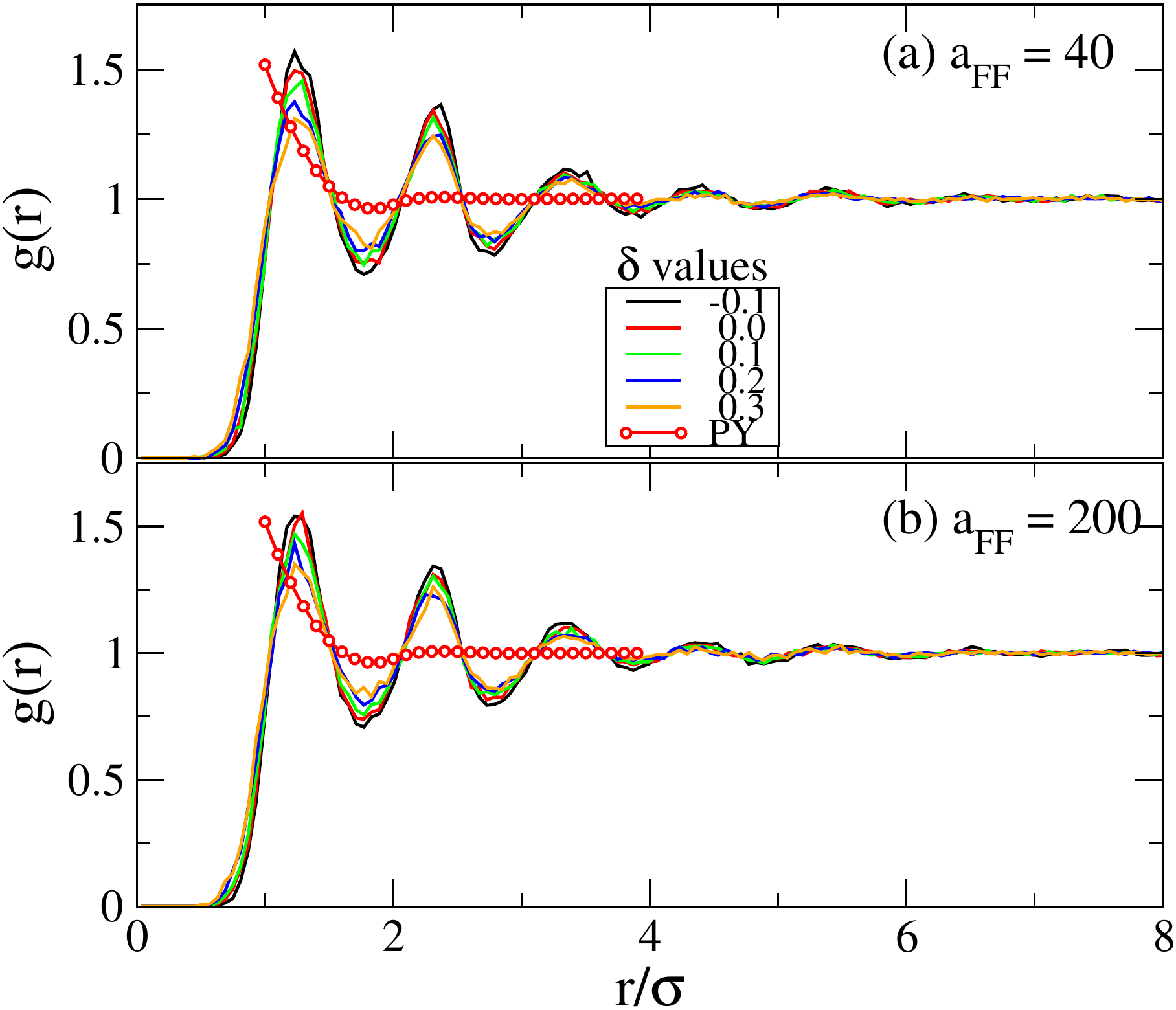} 
		\caption{Radial distribution function $g(r)$ for (a) $a_{\rm FF}=40$ and (b) $a_{\rm FF}=200$ fluid systems shown for $\delta = -0.1~{\rm to}~0.3$ as indicated in the figure. The numerical solution of the Percus-Yevick (PY) equation~\cite{throope1965} for $g(r)$ for the hard sphere potential at packing fraction $\eta = 0.3$ is also shown as reference.} 
		\label{fig: fluid-rdf}
	\end{center}
\end{figure}

To ascertain the systems are indeed in the fluid state we calculate the radial distribution function (which characterizes the spatial organization of the particles) for the systems relaxed at different values of $\delta$ mentioned above by cooling down from high temperature ($T=5$). Mathematically, the radial distributioin function is defined as    
\begin{eqnarray}
g(r) = \frac{V}{N^2} \left< \sum_{i}\sum_{i\neq j}\delta(r - r_{ij})\right> ~,
\label{eqn: rdf}
\end{eqnarray} 
with $r_{ij} = |{\bf r}_i - {\bf r}_j|$, $V$ as volume, and $N$ the number of particles.    
In figure~\ref{fig: fluid-rdf}, we display $g(r)$ curves at different values of $\delta$ together with that of classical hard sphere fluid at the same packing fraction (Percus-Yevick solution~\cite{throope1965}) as a reference. Indeed all the curves show fluid-like behaviour, i.e., oscillates and decays rapidly, also with increasing temperature the amplitude of the curve decreases. In the following we briefly characterize the dynamics of fluid particles which will be followed by the tracer dynamics. 
Dynamics of the fluid particles is characterized via mean-square displacements (MSD) defined as 
\begin{eqnarray}
\langle \Delta r^2(t) \rangle = \frac{1}{N}\sum_{i=1}^N\left[r_i(t) - r_i(0)\right]^2~,
\label{eqn: msd-ens-avg} 
\end{eqnarray}
with $r_i(t)$ the position of $i^{\rm th}$ particle at time $t$. From which the diffusion constant $D_0$ is obtained as 
\begin{eqnarray}
D_0 = \lim_{t\rightarrow \infty} \frac{\langle \Delta r^2(t) \rangle}{6t}.
\label{eqn: diffusion-coeff} 
\end{eqnarray}

\begin{figure}[]
	\begin{center}
		\includegraphics[width=0.45\textwidth]{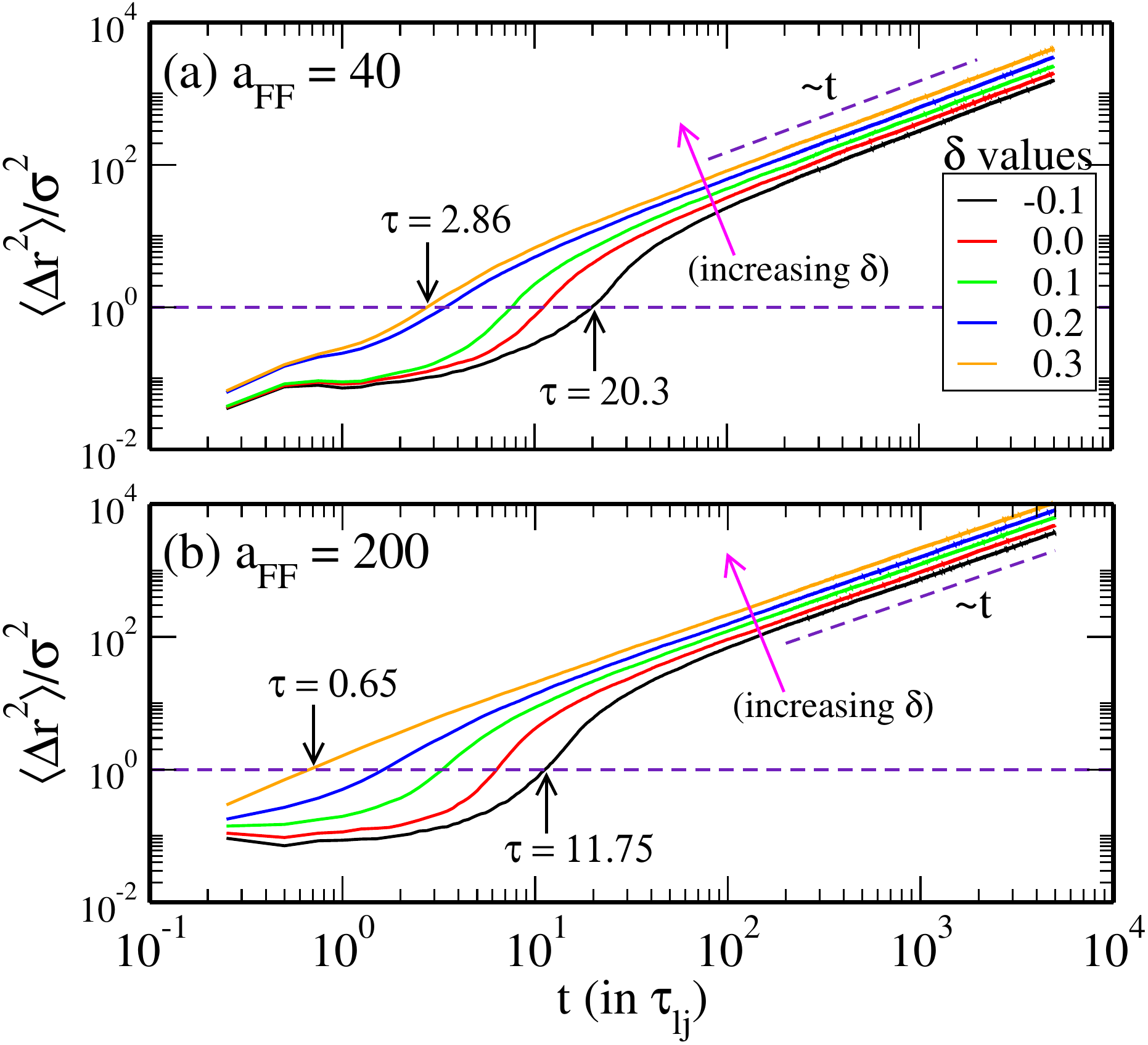} 
		\caption{MSD curves for (a) $a_{\rm FF}=40$ and (b) $a_{\rm FF}=200$ fluid systems shown for $\delta = -0.1~{\rm to}~0.3$ indicated in the figure. The intersection of the horizontal dashed line at $\langle\Delta r^2\rangle/\sigma^2 = 1$ with the MSD curves gives the average relaxation time $\tau$.}
		\label{fig: msd-fluid}
	\end{center}
\end{figure}
\begin{figure}[]
	\begin{center}
		\includegraphics[width=0.5\textwidth]{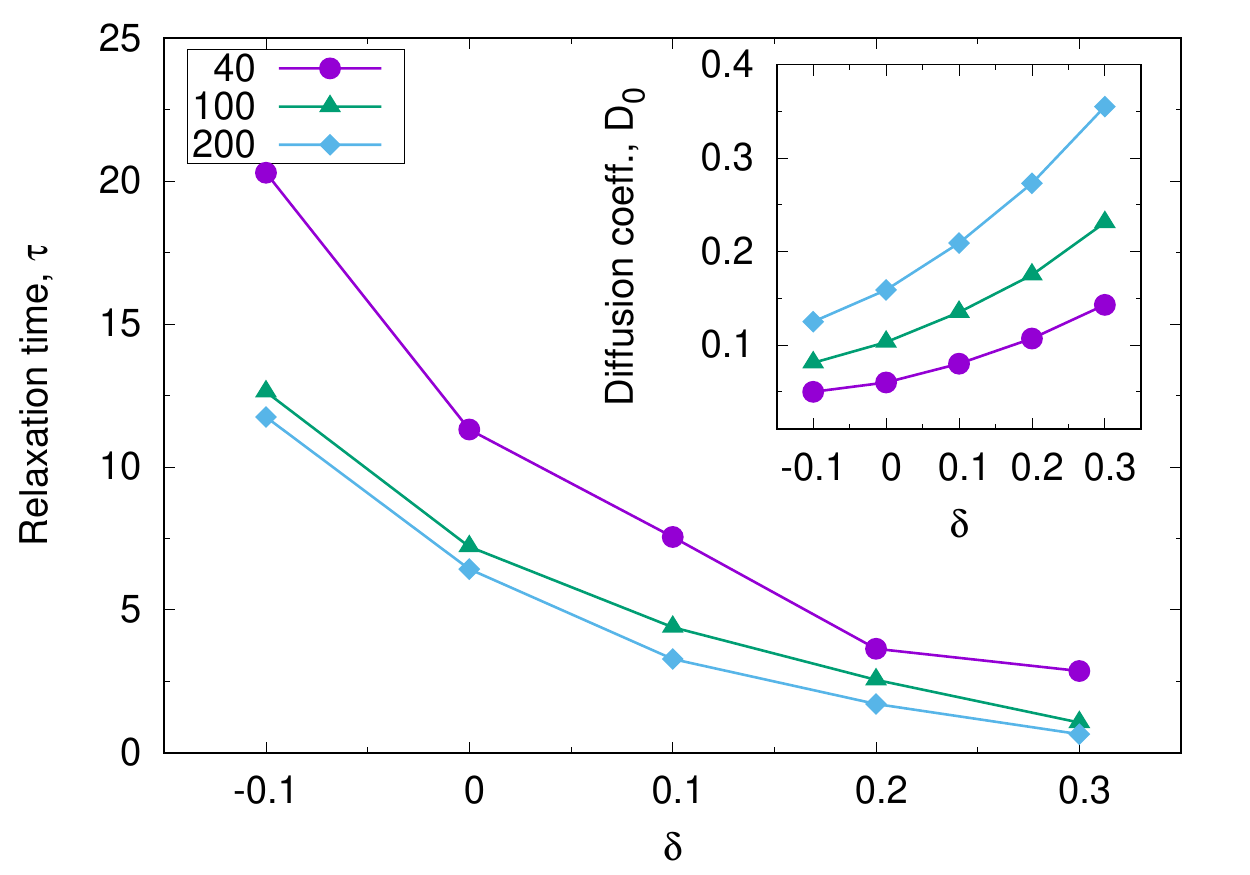} 
		\caption{Mean relaxation time, $\tau$, of the fluid particles plotted as a function of reduced temperature, $\delta$, shown for three different values of fluid particle softness $a_{\rm FF}=40,100,200$ indicated in the figure. Inset figure is the corresponding diffusion coefficient $D_0$ (equation~\ref{eqn: diffusion-coeff}).}
		\label{fig: diffusion-coeff-n-relaxation-time} 
	\end{center}
\end{figure}

In figure~\ref{fig: msd-fluid} we show the MSD curves obtained using equation~\ref{eqn: msd-ens-avg} at different temperatures for the two fluid systems. For all the considered temperature range, it is observed that the MSD curves crosses over to diffusive regime, i.e., $\langle \Delta r^2 \rangle \sim t$ at sufficiently long time scale. The relaxation time together with the diffusion constant $D_0$ at different values of $\delta$ is plotted in figure~\ref{fig: diffusion-coeff-n-relaxation-time}. Here, the relaxation time $\tau$ is defined as the average time for a particle to cover its own size. In the considered colloidal fluid systems, the value of $\tau$ increases (and hence $D_0$ decreases) with lowering $\delta$ as expected since with lowering temperature particle takes longer time to escape the cage formed by the neighbouring particles. In comparison, the value of $\tau$ is relatively large for $a_{\rm FF}=40$ system in the entire range of $\delta$ considered and consequently the value of $D_0$ is consistently lower in the range.  
In the following, we focus on the tracer particles and study the effect of varying softness of colloidal particles on the dynamics of tracer. 



\subsection{Tracer dynamics} \label{sec: dynamics of tracer} 

\begin{figure}[]
	\begin{center}
		\includegraphics[width=0.45\textwidth]{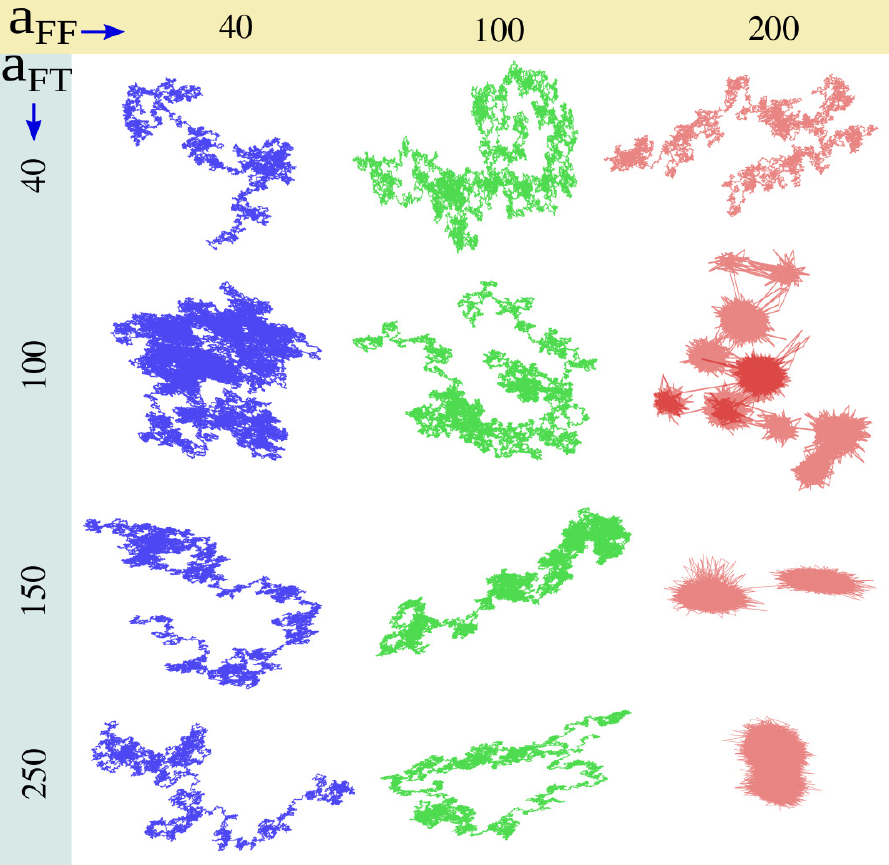} 
		\caption{Typical tracer trajectories (not to scale) obtained at $\delta \approx 0$ for the selected softness parameters $a_{\rm FF}=40,100,200$ and $a_{\rm FT}=40,100,150,250$. All the trajectories shown are of length $1\times 10^7$ simulation steps recorded at the same time interval.}
		\label{fig: tracer-trajectory} 
	\end{center}
\end{figure}

Typical tracer particle trajectories at four different values of fluid-tracer interaction parameter $a_{\rm FT}=$ 40, 100, 150, and 200 are shown in figure~\ref{fig: tracer-trajectory} for the tracers in three different fluid systems characterized by parameter $a_{\rm FF}=$ 40,~100, and~200 at $\delta=0$. We can see that for tracer in ultra-soft fluid particles, i.e. $a_{\rm FF}=40$ in the figure, the trajectories resemble that of normal diffusion, while for tracer in harder fluid particles the motion is intermittent, i.e., it consists of long periods of localized vibrations followed by sudden jumps, see $a_{\rm FF}=200$ and $a_{\rm FT}\ge100$ in the figure. Such intermittent motion is observed in a wide variety of systems that ranges from relaxation in disordered materials to protein trajectories in live cells.\cite{izeddin2014,weeks2000,bouchaud1990,sastry1998} From the visual inspection of the trajectories it is clear that the qualitative feature of the tracer motion depends on the matrix in which it is embedded or the soft/hard-ness of the tracer particle. \\

To gain further insight into the dynamics of tracer in the dense colloidal suspensions under various parameter conditions we first look at the behavior of mean-square displacement (MSD) whence we characterize the relaxation times and diffusion coefficient, and then we investigated the distribution of displacements to understand if any dynamical heterogeniety is visible. 


\subsubsection{Mean-square displacement}
\label{subsec: msd} 

The tracer dynamics is quantified through time averaged MSD (instead of ensemble averaged) as done in single particle tracking experiments where the quantity is extracted from the available position time series ${\bf r}(t)$ of a tracer molecule and then averaged over $N_T~(=10)$ tracer particles. 
Mathematically, the time averaged MSD of a tracer particle is defined as 
\begin{eqnarray}
\overline{\delta^2(\Delta,t)} \equiv \frac{1}{t-\Delta} \int_{0}^{t-\Delta} \left[{\bf r}(t'+\Delta)-{\bf r}(t')\right]^2 dt'~,
\label{eqn: msd-time-avg} 
\end{eqnarray}
where the overline $\overline{(.)}$ denotes the time average. Here, $\Delta$ is the lag time constituting a time window swept along the time series which is less than the total measurement time $t$. The time averaged MSD thus compares the particle positions along the trajectory which is separated by the time difference $\Delta$. Over long-time measurement, the time averaged MSD is related to the diffusion exponent $\alpha$ as 
\begin{eqnarray}
\overline{\delta^2} \sim D_0\Delta^{\alpha}~,
\label{eqn: diffusion-exponent}
\end{eqnarray} 
where $D_0$ is the diffusion constant and the exponent typically lies in the range $0 < \alpha \le 1$ (with $\alpha=1$ for Brownian motion and $\alpha < 1$ for subdiffusive or anomalous diffusion, and for driven systems super-diffusive motion can be observed where $\alpha>1$).\\

\begin{figure*}[]
	\begin{center}
		\includegraphics[width=0.95\textwidth]{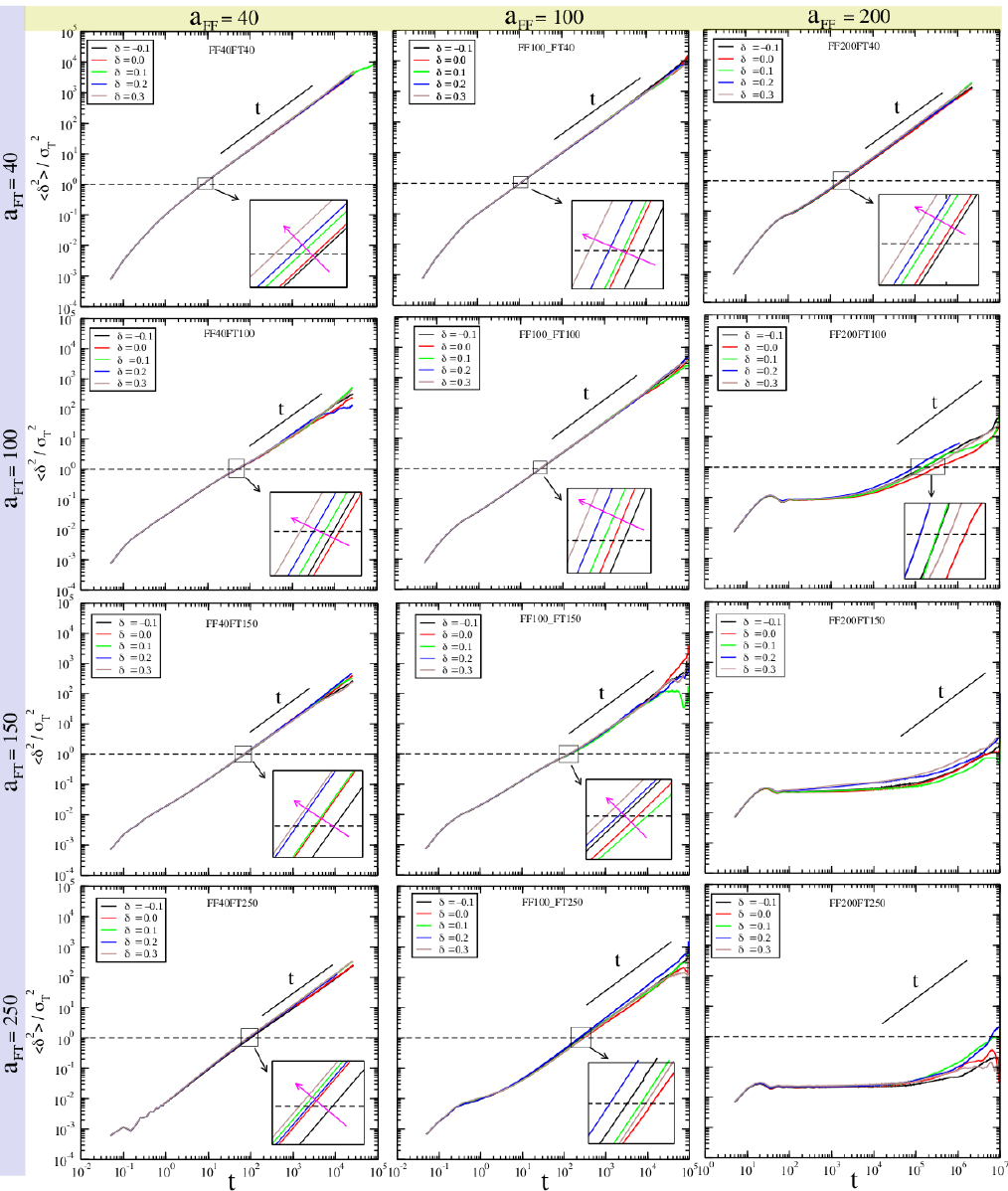} 
		\caption{Time averaged MSD curves at different temperatures ($-0.1 \le \delta \le 0.3$) for the tracer particles obtained at different values of $a_{\rm FF}$ and $a_{\rm FT}$ indicated in the figure. Horizontal dashed-line is drawn at $\langle \delta^2 \rangle /\sigma_{\rm T}^2 = 1$. Inset figure: Zoom in of the encircled region where $\langle \delta^2 \rangle /\sigma_{\rm T}^2 = 1$, and arrow indicates the direction of increasing temperature.}
		\label{fig: msd-tracer-compare} 
	\end{center}
\end{figure*}

The time averaged MSD curves (also averaged over $N_{\rm T}$ tracers) obtained at different values of reduced temperature in the range $-0.1 \le \delta \le 0.3$ (i.e., in the vicinity of $T^\ast$) is shown in figure~\ref{fig: msd-tracer-compare} for different values of softness parameters. Tracer dynamics in the matrix of (ultra-) soft colloids ($a_{\rm FF}=40$ in the figure) is found to be diffusive, i.e.~$\delta^2 \sim \Delta^1$, for all values of $a_{\rm FT}$ in the range $40-250$ considered in this study. And for a given value of $(a_{\rm FF}, a_{\rm FT})$ the particles diffuses faster at higher temperature as expected. However, the average time for the tracer to diffuse its own size (i.e., relaxation time $\tau$) increases with the increase of $a_{\rm FT}$, i.e., harder tracer takes longer time to relax even though the fluid consists of (ultra-) soft colloids. At long time scales, tracers in the fluid systems consisting of relatively hard colloids ($a_{\rm FF}=100,200$ in the figure) show qualitatively similar dynamical behavior, i.e. diffusive. On the other hand, the effect of caging appears (following inertial regime, i.e., $\delta^2 \sim \Delta^2$) at early times and the effect gets stronger with increasing value of $a_{\rm FT}$, as apparent from the appearance of plateau in the MSD curves, and thus the relaxation time shifts to higher value. For example, for $a_{\rm FF}=200$ fluid system, when $a_{\rm FT}=150-200$ the caging effect is very strong and within the reported simulation time scales the tracer hardly covers its own size, see figure~\ref{fig: tracer-trajectory} for the appearance of caging effect with changing softness parameter. It is interesting to note that the range of inertial regime gets extended as tracer gets softer (i.e.~decreasing $a_{\rm FT}$) indicating the fact that due to the soft interaction potential the tracer in a cage are allowed to overlap significantly and thus two particles can approach linearly by considerable amount before reversing its direction. \\

\begin{figure}[]
	\begin{center}
		\includegraphics[width=0.48\textwidth]{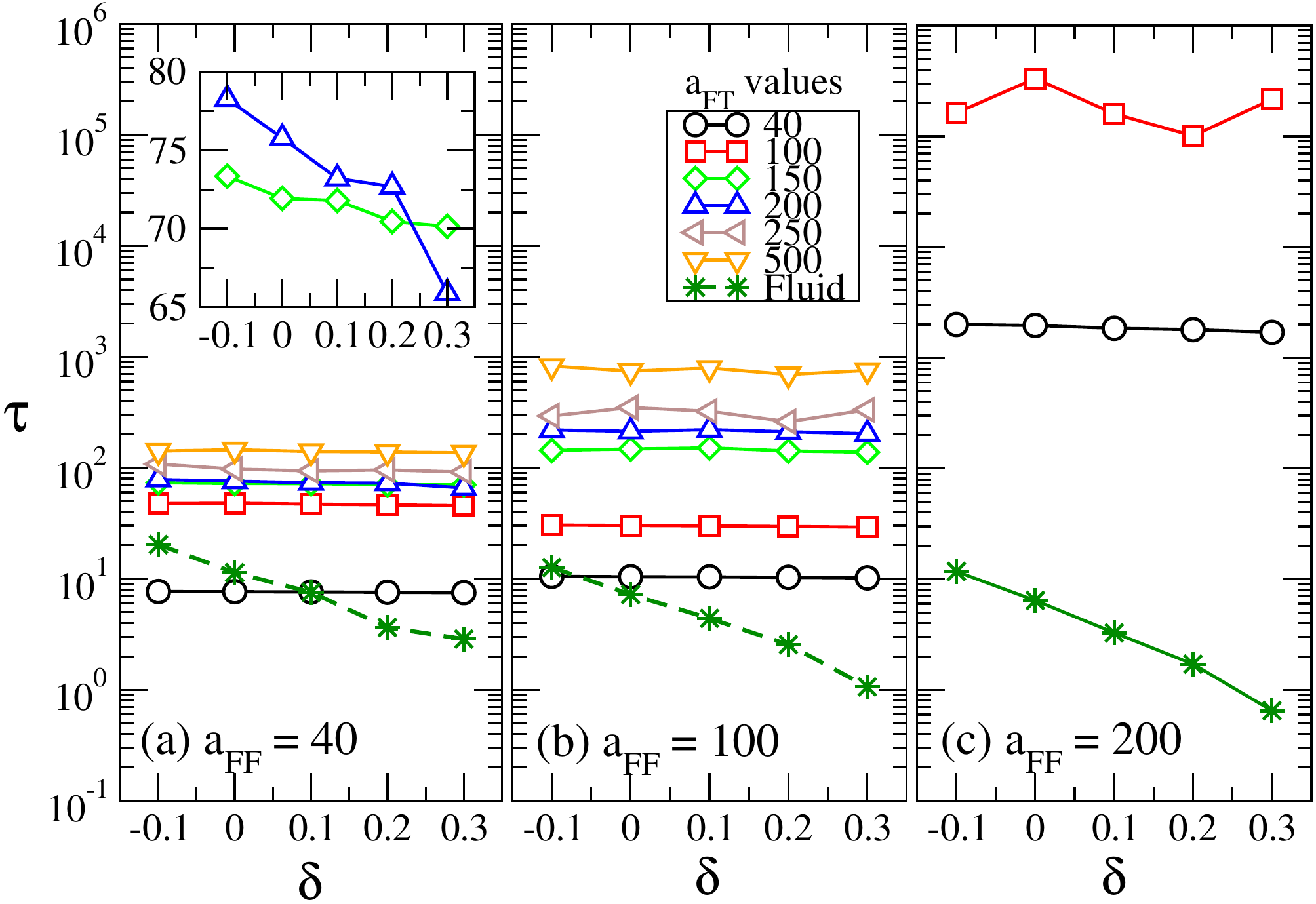} 
		\caption{Relaxation time, $\tau$, of the tracer particles as a function of reduced temperature, $\delta$, shown for different values fluid-tracer interaction parameter, $a_{\rm FT}$, at (a) $a_{\rm FF}=40$, (b) $a_{\rm FF}=100$, and (c) $a_{\rm FF}=200$. For comparison, $\tau$ for the fluid particles of the respective system is also plotted. Inset figure in (a): Linear scale plot for two selected values of $a_{\rm FT}$ to show the decrease of $\tau$ with the increase of temperature (which is not clearly visible in the log scale).} 
		\label{fig: relaxation-time-compare} 
	\end{center}
\end{figure}

The relaxation time $\tau$ of the tracer particle as a function of temperature at different values of $a_{\rm FT}$ are shown in figure~\ref{fig: relaxation-time-compare} for tracer in the considered three different fluid systems ($a_{\rm FF}=40,100,200$). In general, $\tau$ decreases with increasing $T$ for both tracer and fluid, see figure~\ref{fig: diffusion-coeff-n-relaxation-time} for fluid particles, in all the systems. At a given $T$, $\tau$ for colloids decreases with increasing $a_{\rm FF}$ (or increasing hardness of colloids). On the contrary, $\tau$ increases for tracer with increasing $a_{\rm FT}$, i.e., as the tracer gets harder it takes longer time to relax. Depending on the softness of the colloidal suspension the value of $\tau$ can vary by an order of magnitude or more with changing $a_{\rm FT}$, see figure~\ref{fig: relaxation-time-compare}(a)-(b) for ultra and intermediate degree of softness, and for tracer in relatively hard colloidal suspension it is trapped and thus could not relax (within the simulation time scale reported here) for $a_{\rm FT}>100$, see figure~\ref{fig: relaxation-time-compare}(c). Because of its bigger size the relaxation time of tracer, in general, is expected to be large compared to colloids. However, for $a_{\rm FF}=a_{\rm FT}=40$ system, it is interesting to note that $\tau_{\rm tracer}$ is comparable to $\tau_{\rm colloid}$, see figure~\ref{fig: relaxation-time-compare}(a). \\

\begin{figure}[]
	\begin{center}
		\includegraphics[width=0.45\textwidth]{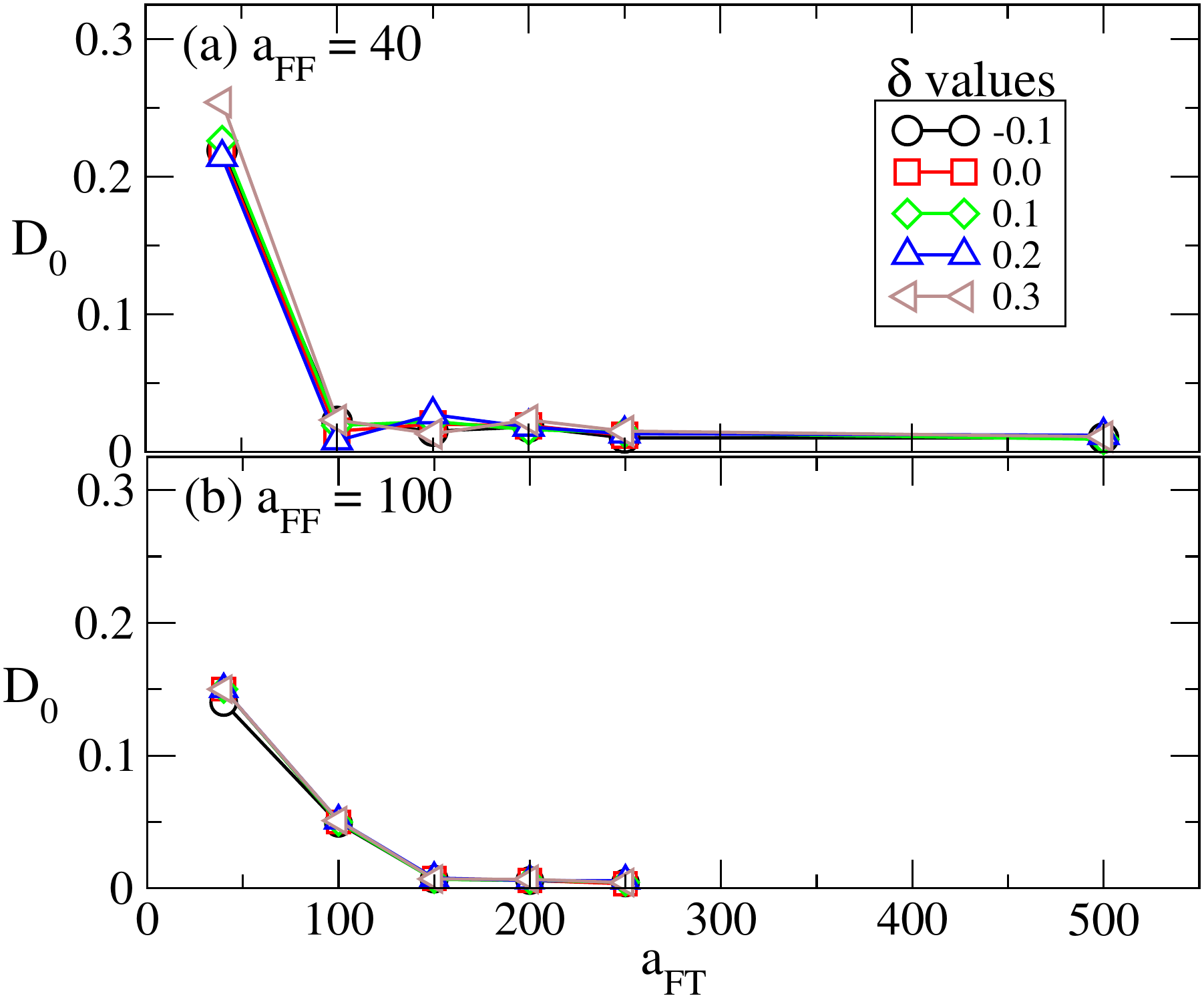} 
		\caption{Diffusion constant, $D_0$, of the tracer particles as a function of softness parameter $a_{\rm FT}$ for tracer in (a) $a_{\rm FF}=40$, and (b) $a_{\rm FF}=100$ colloidal suspensions.} 
		\label{fig: DiffCoeff-compare} 
	\end{center}
\end{figure}

As shown in figure~\ref{fig: DiffCoeff-compare}, diffusion constant $D_0$ is calculated for tracer in colloidal suspensions with $a_{\rm FF}=$ 40 and 100 where the systems exhibiting diffusive motion for sufficient time (at least three decades for all the values of $a_{\rm FT}$ considered) within the simulation time scale. (In fact, for tracer in $a_{\rm FF}=200$ colloidal suspension, $a_{\rm FT}=$ 40 and 100  systems also show diffusion for at least two decades in time.) It is observed that for tracer in ultra-soft colloidal suspension, when $a_{\rm FT}=40$ we get $D_0$ in the range $0.22-0.25$ depending on the temperature which is dropped by a factor of about 10 upon increasing $a_{\rm FT}$ to 100 and change thereafter is relatively very small, see figure~\ref{fig: DiffCoeff-compare}(a). Whereas, in the case of tracer in $a_{\rm FF}=100$ colloidal suspenson, $D_0 \approx 0.15$ and the decrease of $D_0$ with increasing $a_{\rm FT}$ is rather gradual. 
Since MSD curves represent an averaged over all the possible values of displacements that the tracer made in a given time interval we cannot see even if distinct dynamical regimes are present. Therefore, in order to gain deeper insights we now proceed to calculating the distribution of displacements described in the following section.  


\subsubsection{Displacement distribution} \label{subsec: displ-distribution} 

In many systems, e.g. glass formers or supercooled liquid,\cite{sastry1998} distinct dynamical regimes are observed closed to criticality. The qualitative change in the dynamics or dynamical heterogeneity close to the transition is well-captured by the displacement distribution also known as van Hove self-correlation function defined as
\begin{eqnarray}
G_s(r,\Delta) = \langle \delta(r-|r_i(\Delta)-r_i(0)|)  \rangle ~,
\label{eqn: displacement-dist}
\end{eqnarray}
where $\Delta$ is the time lag and $r_i(\Delta)$ is the position of $i^{\rm th}$ tracer at time lag  $\Delta$ along the trajectory. Here, $\delta(\cdot)$ represents the Dirac delta function. The quantity $G_s(r,\Delta)$ describes the probability of finding a particle at distance $r$ from a given initial position after time interval $\Delta$ along the particle trajectory. Here, all the particles positions are measured with respect to the instantaneous center of mass of the system, and we focus our attention to the dynamics near the transition temperature where the differences in the trajectories is expected to magnify.\\ 
 
\begin{figure*}[]
	\begin{center}
		\includegraphics[width=0.95\textwidth]{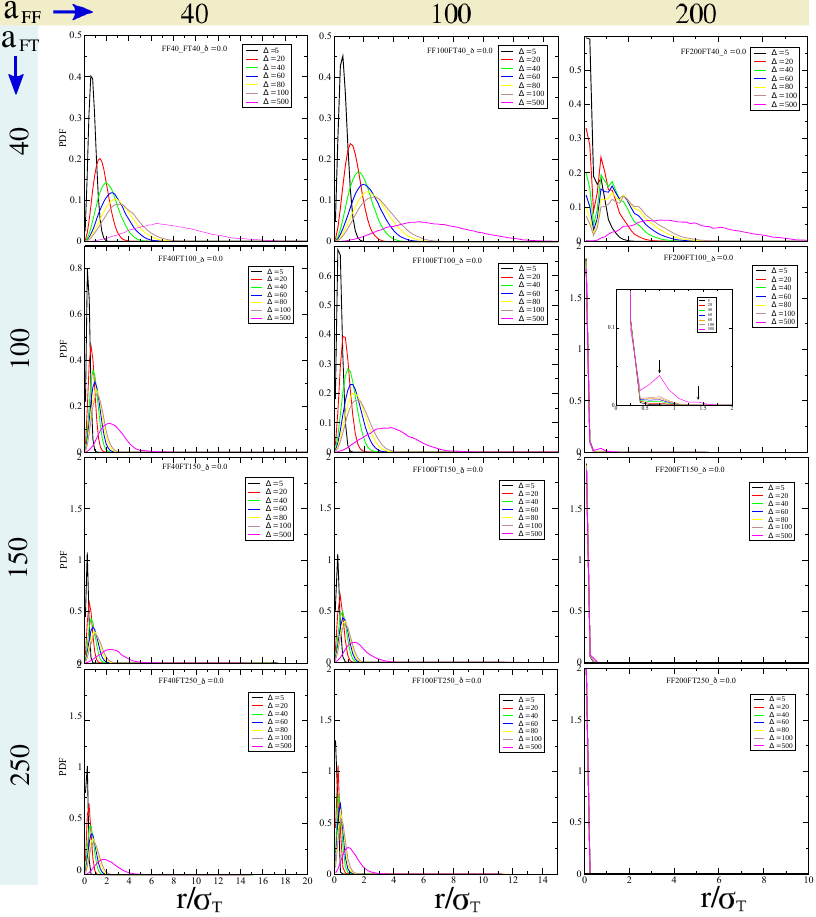} 
		\caption{Displacement distributions obtained at different values of time-lag $\Delta$ in the range $5-500$ shown for fluid-tracer softness parameter $a_{\rm FT}=40,100,150,250$ (top to bottom) for tracer in the three different fluid systems characterized by softness parameter $a_{\rm FF}=40,100,200$ (left to right).} 
		\label{fig: displ-histo} 
	\end{center}
\end{figure*}

The displacement distributions obtained at different values of $\Delta$ in the range $5-500$ are shown in figure~\ref{fig: displ-histo} for different values of softness parameters $a_{\rm FF}$ and $a_{\rm FT}$. The displacement distribution is Gaussian for tracer in ultra-soft colloidal suspensions ($a_{\rm FF}=40$ in the figure) whose width gets broader with increasing $\Delta$ for a given value of $a_{\rm FT}$, and as $a_{\rm FT}$ increases the range of the distribution decreases. This observation is consistent with the fact that the tracer motion is diffusive $\overline{\delta^2}\sim \Delta^1$ and the structural relaxation time increases with increasing $a_{\rm FT}$. Similar behaviour is observed for tracer in fluid system with $a_{\rm FF}=100$.\\

In general, the broadening of the distributions with time indicates that the tracer eventually escapes the cage. Now, for tracer in the relative hard colloidal suspensions ($a_{\rm FF}=200$ in the figure) the distributions show the following behavior: (i) at $a_{\rm FT}=40$, in addition to the first order peak (corresponding to the vibration about the mean position), higher order peaks are observed at $r/\sigma_{\rm T}\approx 0.75,~1.3,~1.75$ (for $\Delta < 500$) although the overall curves looks Gaussian which is consistent with the earlier observation that $\overline{\delta^2}\sim \Delta^1$ in this case, (ii) at $a_{\rm FT}=100$ higher order peaks around $r/\sigma_{\rm T}\approx 0.75,~1.3$ are observed for $\Delta = 500$ which is in agreement with the visual inspection of the trajectory shown in figure~\ref{fig: tracer-trajectory}, (iii) $a_{\rm FT} > 100$ tracers are strongly trapped in the cage formed by the neighboring colloids and thus a sharp narrow peak corresponding to the vibration about the mean position is only prominent within the time scale under consideration.    




\section{Conclusions} \label{sec: conclusion} 

In this study, we have investigated the dynamics of tracer particles embeded in dense colloidal suspensions (in fluid phase and close to the thermodynamic freezing temperature where the effect of pair-wise interaction becomes significant) by means of DPD simulations. For relatively large tracer particles (i.e., three times larger than the colloid) we looked at the effect on the dynamics due to variation in the fluid-tracer interaction from being ultra-soft to hard (mimicking highly deformable or interpenetrable to rigid tracer particles). Relation between tracer dynamics and underlying fluid medium is also systematically explored by considering three different colloidal suspensions (or fluid systems) differing in their degree of softness. Since the fluid systems have different values of thermodynamic melting or freezing temperature comparison among the systems are done at the same value of reduced temperature $\delta$ which represents the relative distance from the respective transition temperature, $T^*$. At a given value of $T$, the dynamics of the fluid particles is sensitive to its softness, e.g. diffusion constant (or relaxation time) increases (or decreases) as particle gets harder, which in turn affect the dynamics of tracer particles. \\ 

It was observed that for tracer in ultra-soft and intermediate hardness colloidal suspensions its motion is diffusive at long times (for soft/hard tracer particles considered in this study) and the relaxation time increases with increasing hardness of tracer particles which in turn reduces the diffusion constant. However, a qualitatively different tracer motion is observed for tracer in hard coolidal suspension where the motion changes from that of a free diffusion to intermittent jumps after a long period of localized vibrations - resembles hopping motion. Strong caging where tracer particles escapes from the cage after a very long time is also observed as tracer particles get harder which is highlighted by single large rattling in the trajectories. Furthermore, a closer look at the dynamics through displacement distribution reveal that for tracer exhibiting free diffusion process (i.e. for tracer in ultra-soft and intermediate colloidal suspensions) the distribution is Gaussian whose width decreases as tracer gets harder. Interestingly, for tracer showing hopping or discrete type motion higher order peaks are observed in the distribution reflecting the different qualitative behavior at different time scales. Despite the fact that the model considered in this study is far simple than the real systems it clearly illustrates the interplay between the medium in which tracer undergoes motion and the nature of colloid-tracer interaction. We shall explore in a later work how the shape of tracer affects its dynamics in this dense soft-colloidal suspensions. 


\begin{acknowledgements}
J. Samukcham acknowledges fruitful discussions with M.~Premjit, J.~Pame, T.~Vilip, T.~Premkumar, and U.~Somas.
\end{acknowledgements}

\end{document}